\documentclass[%
reprint,
superscriptaddress,
 amsmath,amssymb,
 aps,
]{revtex4-2}

\usepackage{graphicx}
\usepackage{dcolumn}
\usepackage{bm}
\usepackage{nicefrac}
\usepackage{multirow,wrapfig,array,tabularx,rotating,verbatim,float,mathtools,booktabs,color,physics}
\definecolor{Blue}{rgb}{0.24,0.24,0.76}

\begin{document}


\title{Electronic structure and optical properties of Na$_2$KSb and NaK$_2$Sb from first-principles many-body theory}

\author{Raymond Amador}
\affiliation{%
 Humboldt-Universit{\"a}t zu Berlin, Physics Department and IRIS Adlershof, 12489 Berlin, Germany
}%
\affiliation{%
 Helmholtz-Zentrum Berlin, 12489, Berlin, Germany
}%

\author{Holger-Dietrich Sa{\ss}nick}%
\affiliation{%
Carl von Ossietzky Universit\"at Oldenburg, Institute of Physics, 26129 Oldenburg, Germany
}%

\author{Caterina Cocchi}%
 \email{caterina.cocchi@uni-oldenburg.de}
\affiliation{%
 Humboldt-Universit{\"a}t zu Berlin, Physics Department and IRIS Adlershof, 12489 Berlin, Germany
}%
\affiliation{%
Carl von Ossietzky Universit\"at Oldenburg, Institute of Physics, 26129 Oldenburg, Germany
}%

\date{\today}%

\begin{abstract}
In the search for novel materials for vacuum electron sources, multi-alkali antimonides and in particular sodium-potassium-antimonides have been recently regarded as especially promising due to their favorable electronic and optical properties.
In the framework of density-functional theory and many-body perturbation theory, we investigate the electronic structure and the dielectric response of two representative members of this family, namely Na$_2$KSb and NaK$_2$Sb. 
We find that both materials have a direct gap, which is on the order of 1.5~eV in Na$_2$KSb and 1.0~eV  in NaK$_2$Sb.  
In either system, valence and conduction bands are dominated by Sb states with $p$- and $s$-character, respectively.
The imaginary part of the dielectric function, computed upon explicit inclusion of electron-hole interactions to characterize the optical response of the materials, exhibits maxima starting from the near-infrared region, extending up to the visible and the ultraviolet band.
With our analysis, we clarify that the lowest-energy excitations are non-excitonic in nature and that their binding energy is on the order of 100~meV.
Our results confirm the potential of Na$_2$KSb and NaK$_2$Sb as photoemissive materials for vacuum electron sources, photomultipliers, and imaging devices. 
\end{abstract}

\maketitle


\section{Introduction}
Multi-alkali antimonides comprise an interesting class of semiconducting compounds that have become popular for use as vacuum electron sources at particle accelerators, due to their favorable electronic properties~\cite{hern+08pt,dowe+10nimpra,musu+18nimpra}.
Compared to conventional metals such as copper, which have been employed for decades to build photocathodes, alkali antimonides offer several advantages that are key for the development of the next generation of electron sources~\cite{musu+18nimpra}.
Their absorption onset in the visible region and their photoemission threshold at infrared frequencies ideally match the operational conditions of photocathodes illuminated by green laser pulses~\cite{musu+18nimpra,schm+18prab}.
These characteristics prevent inefficient energy conversion processes and together with the intrinsically low thermal emittance exhibited by semiconductors maximize the quantum efficiency of the cathode~\cite{dunh+13apl}, which, posed in simple terms, represents the amount of emitted photoelectrons with respect to the amount of incident photons.
Recent efforts of several groups worldwide have demonstrated the superiority of multi-alkali antimonides in terms of both quantum efficiency and mean transverse emittance (roughly speaking, the size of the emitted electron beam)~\cite{mich+94nimpra,dibo+97nimpra,smed+09aipcp,cult+11apl,vecc+11apl,schu+13aplm,mamm+13prab,ruiz+14aplm,cult+15prab,xie+16prab,schm+18prab} also in comparison with more widely studied semiconductors like GaAs~\cite{orlo+04nimpra,sinc+07prab,kuri+11nimpra}. 

From the experimental perspective, the routinely adopted deposition methods~\cite{maxs+15apl,ding+17jap,panu+21nimpra}, together with the propensity of multi-alkali antimonides to be contaminated by atmospheric pollutants~\cite{schu+13aplm}, demand ultrahigh vacuum conditions for both growth and characterization~\cite{erja01vacuum,cult+16jvstb} which minimize opportunities for thorough analysis.
In light of these limitations, results of \textit{ab initio} calculations represent a valuable source to obtain information about the properties of these systems. 
The pioneering studies by Ettema and de Groot~\cite{ette-degr00prb,ette-degr02prb}, based on the localized spherical wave method, demonstrated the potential of multi-alkali antimonides as photocathode materials.
More recently, a few works based on density-functional theory (DFT) provided additional references to the electronic structure of this family of compounds~\cite{kala+10jpcs,kala+10jpcs1,yala+18jmmm,khan+20ijer}.
An important step forward, going beyond the mean-field picture of DFT, was offered by recent studies on Cs-based multi-alkali antimonides~\cite{cocc+18jpcm,cocc+19sr}. 
The explicit inclusion of electron-electron and electron-hole interactions in the framework of many-body perturbation theory (MBPT) on top of DFT improves the accuracy in the description of the microscopic characteristics of these materials.
Quantitative estimates of the fundamental and optical gaps and insight regarding the nature and composition of the lowest-energy excitations are essential quantities to subsequently evaluate the performance of a photocahode.
To this end, the three-step model for photoemission, originally developed by Berglund and Spicer~\cite{berg-spic64pr}, is still largely applied and the very recent generalization to embed DFT results~\cite{anto+2020prb} opens promising perspectives to predict operational characteristics of photocathodes based on microscopic properties computed from first principles.

Research on multi-alkali antimonides has been so far mainly devoted to Cs-based compounds~\cite{smed+09aipcp,vecc+11apl,baza+11apl,cult+15prab,xie+16prab,schm+18prab}.
However, interest is growing also towards the Na-based members of this family~\cite{nata+01jap,cult+13apl,maxs+15apl,cult+16jvstb,cult+16apl}, reviving the predictions of early studies~\cite{chik+61jpsj,fish+74jap,doli+88}.
The higher electronegativity and lower atomic weight of sodium compared to heavier alkali metals such as cesium is expected to give rise to more advantageous characteristics for photocathode applications. 
In particular, their tolerance for high temperatures up to 200$^{\circ}$ and lower dark current is particularly appealing in particle accelerators.
However, the realization of this high potential is hindered by the current lack of systematic studies, both theoretical and experimental, which can unravel and, most importantly, rationalize the microscopic physical properties of Na-based multi-alkali antimonides.

With this \textit{ab initio} study based on DFT and MBPT, we aim to characterize the electronic and optical properties of Na$_2$KSb and NaK$_2$Sb, in order to provide reliable references for the development of Na-based antimonides as photocathode materials.
To this end, we focus our analysis on the band structures of the systems, unraveling nature and size of the fundamental gaps, and inspecting the character of the valence and conduction states that are prominently involved in the photoabsorption and emission processes.
Moreover, from the calculation of the dielectric function including explicitly electron-electron and electron-hole interactions, we gain insight into the energies and composition of the optical excitations.

This paper is organized as follows. 
In Section~\ref{sec:theory}, we review the adopted DFT and MBPT formalism and provide the relevant computational details.
In Section~\ref{sec:results}, the body of results are presented, including structural properties of Na$_2$KSb and NaK$_2$Sb (Section~\ref{subsec:struct}), electronic structure (Section~\ref{subsec:el-struct}), and optical excitations (Section~\ref{subsec:opt}).
In Section~\ref{sec:conclu} we summarize our findings and report our conclusions.

\section{\label{sec:theory}Theoretical Background and Computational Details}

The results presented in this work are performed in the framework of DFT~\cite{hohe-kohn64pr} and MBPT including the $GW$ approximation and the solution of the BSE~\cite{onid+02rmp}.
The main task of DFT is the solution of the Kohn-Sham (KS) equations~\cite{kohn-sham65pr}, where the accuracy of the results depends on the chosen approximation for the exchange-correlation (xc) potential.

In order to obtain a reliable description of the excited-state properties of the materials,  MBPT is applied on top of the DFT band structure.
In the $GW$ approximation~\cite{hedi65pr}, the quasi-particle (QP) equation is solved to obtain electronic energies including the self-energy contribution.
For this purpose, the single-shot $G_0W_0$ approach~\cite{hybe-loui85prl} is adopted here.
Optical excitations are calculated on top of the QP band structure from the solution of the BSE~\cite{salp-beth51pr}.
In practice, this problem is mapped into the eigenvalue equation
\begin{equation}\label{bsehamiltonian}
\sum\limits_{v'c'\vb{k}'}\hat{H}^{\text{BSE}}_{vc\vb{k},v'c'\vb{k}'}A^\lambda_{v'c'\vb{k}'}=E^\lambda A^\lambda_{vc\vb{k}},
\end{equation}
where for a spin-degenerate system the BSE Hamiltonian $\hat{H}^{\text{BSE}}$ is expressed as $\hat{H}^{BSE} = \hat{H}^{diag} + \hat{H}^{dir} + 2 \hat{H}^x$: The diagonal term, $\hat{H}^{diag}$, accounts for the transition energies between QP states; the direct Coulomb integral, $\hat{H}^{dir}$, includes the screened electron-hole interaction; the exchange term, $\hat{H}^x$, incorporates the short-range repulsive exchange interaction.
For further details about the formalism we refer the readers to specialized reviews~\cite{onid+02rmp,vorw+19es}.

The eigenvalues of Eq.~\eqref{bsehamiltonian} correspond to the excitation energies while the eigenvectors $A^\lambda$ provide information about the character of the excited states and their composition in terms of single-particle transitions. 
They enter the expression of the imaginary part of the macroscopic dielectric function
\begin{equation}
\imaginary\varepsilon_M=\frac{8\pi^2}{\Omega}\sum\limits_\lambda\abs{t^\lambda}^2\delta(\omega-E^\lambda),
\label{eq:Im}
\end{equation}
where $\Omega$ is the unit cell volume and $\omega$ the frequency of the incoming photon, through the transition coefficients
\begin{equation}\label{eq:t}
t^\lambda=\sum\limits_{vc\vb{k}}A^\lambda_{vc\vb{k}}\frac{\mel{v\vb{k}}{\hat{\vb{p}}}{c\vb{k}}}{\epsilon^{QP}_{c\vb{k}}-\epsilon^{QP}_{v\vb{k}}}.
\end{equation}
Eq.~\eqref{eq:Im} describes the optical response of the system.
In cubic crystals, off-diagonal components are zero and the diagonal ones all bear the same value.
In hexagonal crystals, two diagonal components are mutually equal and different from the third one.
Macroscopic quantities, such as absorbance or reflectance can be obtained from the dielectric tensor~\cite{fox02book}, even in case of complex anisotropic crystals~\cite{yeh80ss,pusc-ambr06aem,vorw+16cpc,pass-paar17josab}.

Calculations are performed with the all-electron full-potential code \texttt{exciting}~\cite{gula+14jpcm}, which implements the linearised augmented plane wave plus local orbital formalism. 
This code offers a state-of-the-art implementation of MBPT, treating optical and core spectroscopy on equal footing~\cite{vorw+17prb,vorw+19es}.
The muffin-tin radii of Na and K atoms are set to 2.0~bohr, while the radius of Sb to 2.2~bohr.
A plane-wave basis-set cutoff $R_{\textrm{MT}}G_{\textrm{max}}$~=~8.0 is adopted for cubic Na$_2$KSb, while for hexagonal NaK$_2$Sb this value is increased to 8.5.
In the DFT calculations, the generalized-gradient approximation in the Perdew-Burke-Ernzerhof (PBE) parameterisation~\cite{pbe} is employed for $v_{xc}$.
The Brillouin zone (BZ) is sampled by a 10$\times$10$\times$10 \textbf{k}-mesh in cubic Na$_2$KSb.
In the hexagonal NaK$_2$Sb, we use a 8$\times$8$\times$4 \textbf{k}-grid.

$GW$ calculations~\cite{nabo+16prb} are performed in the single-shot perturbative approach $G_0W_0$ on top of the KS electronic structure.
A 6$\times$6$\times$6 (4$\times$4$\times$2) \textbf{k}-mesh is adopted in cubic Na$_2$KSb (hexagonal NaK$_2$Sb) to sample the BZ at this stage.
Screening is evaluated in the random-phase approximation (RPA) including 200 empty states in both systems, with the self-energy computed by analytic continuation.
BSE calculations are performed with the Tamm-Dancoff approximation on a $\Gamma$-shifted \textbf{k}-mesh with 8$\times$8$\times$8 points and 4$\times$4$\times$2 for cubic Na$_2$KSb and hexagonal NaK$_2$Sb, respectively. 
The screened Coulomb interaction is computed within the RPA using 100 empty states in both cases.
Local-field effects are included with a cutoff  of 1.5~Ha.
The considered transition space includes 3 (5) valence states and 10 (12) unoccupied bands in cubic Na$_2$KSb (hexagonal NaK$_2$Sb).
These computational parameters ensure an accuracy of the MBPT calculations of the order of 100~meV.

\section{\label{sec:results}Results}

\subsection{Structural properties}
\label{subsec:struct}

\begin{figure}
	\centering
	\includegraphics[width=0.5\textwidth]{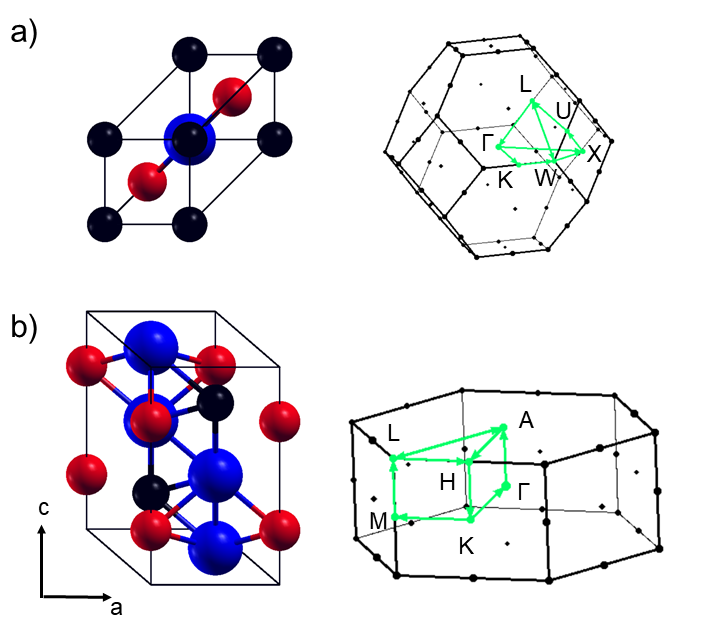}
	\caption{Unit cells (left) and Brillouin zones (right) of a) cubic Na$_2$KSb and b) hexagonal NaK$_2$Sb. In the former, Sb atoms are depicted in black, Na in red, and K in blue. In the latter, the high-symmetry points and the path connecting them used in the calculated band structures are marked. Graphs produced with \texttt{XCrysDen}~\cite{xcrysden}.}
	\label{fig:UnitCellBZ}
\end{figure}

In this work, we consider the two trialkali antimonides with chemical formulas $\text{Na}_2\text{KSb}$ and $\text{NaK}_2\text{Sb}$.
According to the pioneering experimental study performed by McCarroll~\cite{mcca60jpcs}, the former compound is more stable than the latter, which, however, can be formed as an alloy of the type  $\text{Na}_{3-x}\text{K}_x\text{Sb}$ upon a proper combination of the constituting atoms. 
$\text{Na}_2\text{KSb}$ has a face-centered cubic (FCC) Bravais lattice with the Sb atom located at the origin $(0,0,0)$, while the alkali atoms are found at $(\nicefrac{1}{2},\nicefrac{1}{2},\nicefrac{1}{2})$ (K) and $\pm(\nicefrac{1}{4},\nicefrac{1}{4},\nicefrac{1}{4})$ (Na) positions.
$\text{NaK}_2\text{Sb}$ has a hexagonal crystal structure with 8 atoms in the unit cell~\cite{mcca60jpcs}.
To model this structure we adopt the experimentally proposed lattice with two Na atoms at $(0,0,\nicefrac{1}{2}\pm\nicefrac{1}{4})$, four K atoms at $(\nicefrac{1}{3},\nicefrac{2}{3},\nicefrac{3}{4}\pm\nicefrac{1}{6})$ and $(\nicefrac{2}{3},\nicefrac{1}{3},\nicefrac{1}{4}\pm\nicefrac{1}{6})$, and two Sb atoms at $(\nicefrac{1}{2}\pm\nicefrac{1}{6},\nicefrac{1}{2}\mp\nicefrac{1}{6},\nicefrac{1}{2}\mp\nicefrac{1}{4})$.
Starting from the experimental crystal structures~\cite{mcca60jpcs}, a volume optimization based on the Birch-Murnaghan fit~\cite{birch,murnaghan} is performed.
The two structures are depicted in Fig.~\ref{fig:UnitCellBZ}, where also the corresponding Brillouin zones and the high-symmetry points in the conventional notation proposed in Ref.~\cite{sety-curt10cms} are shown.
The lattice parameter obtained for $\text{Na}_2\text{KSb}$ is $a=7.74$~\AA{}, which is 0.02~\AA{} larger than the experimental one~\cite{mcca60jpcs}.
For the hexagonal structure of $\text{NaK}_2\text{Sb}$, we obtain an in-plane lattice parameter $a=5.58$~\AA{} and an out-of-plane lattice vector $c=10.90$~\AA{}.
Both values are 0.03~\AA{} smaller compared to the measurements, which, however, were performed on a non perfectly stoichiometric structure~\cite{mcca60jpcs}.
The $c/a$ ratio equal to 1.95 is in excellent agreement with the experimental reference~\cite{mcca60jpcs}.

The interatomic distances obtained in both compounds are in overall agreement with the experimental values~\cite{mcca60jpcs}. 
We notice that the Na-Sb spacing increases from 3.22~\AA{} to 3.35~\AA{} from the cubic to the hexagonal phase. 
Also the K-Sb distance is subject to a similar behavior being 3.87~\AA{} in $\text{Na}_2\text{KSb}$ and 3.63~\AA{} as well as 3.70~\AA{} in $\text{NaK}_2\text{Sb}$, where two types of K-Sb coordination exist.
Very different trends are instead obtained for the Na-Na and the K-K distances, which are more crucially affected by the crystal structure. 
The former is 5.58~\AA{} in $\text{NaK}_2\text{Sb}$ and 3.87~\AA{} in $\text{Na}_2\text{KSb}$, while, conversely, the latter is 3.63~\AA{} in the hexagonal phase and 5.47~\AA{} in the cubic one. 
Finally, the Na-K distance is larger in $\text{NaK}_2\text{Sb}$ (3.70~\AA{}) than in  $\text{Na}_2\text{KSb}$ (3.35~\AA{}).

\subsection{Electronic structure}
\label{subsec:el-struct}

\begin{figure}
	\centering
	\includegraphics[width=0.5\textwidth]{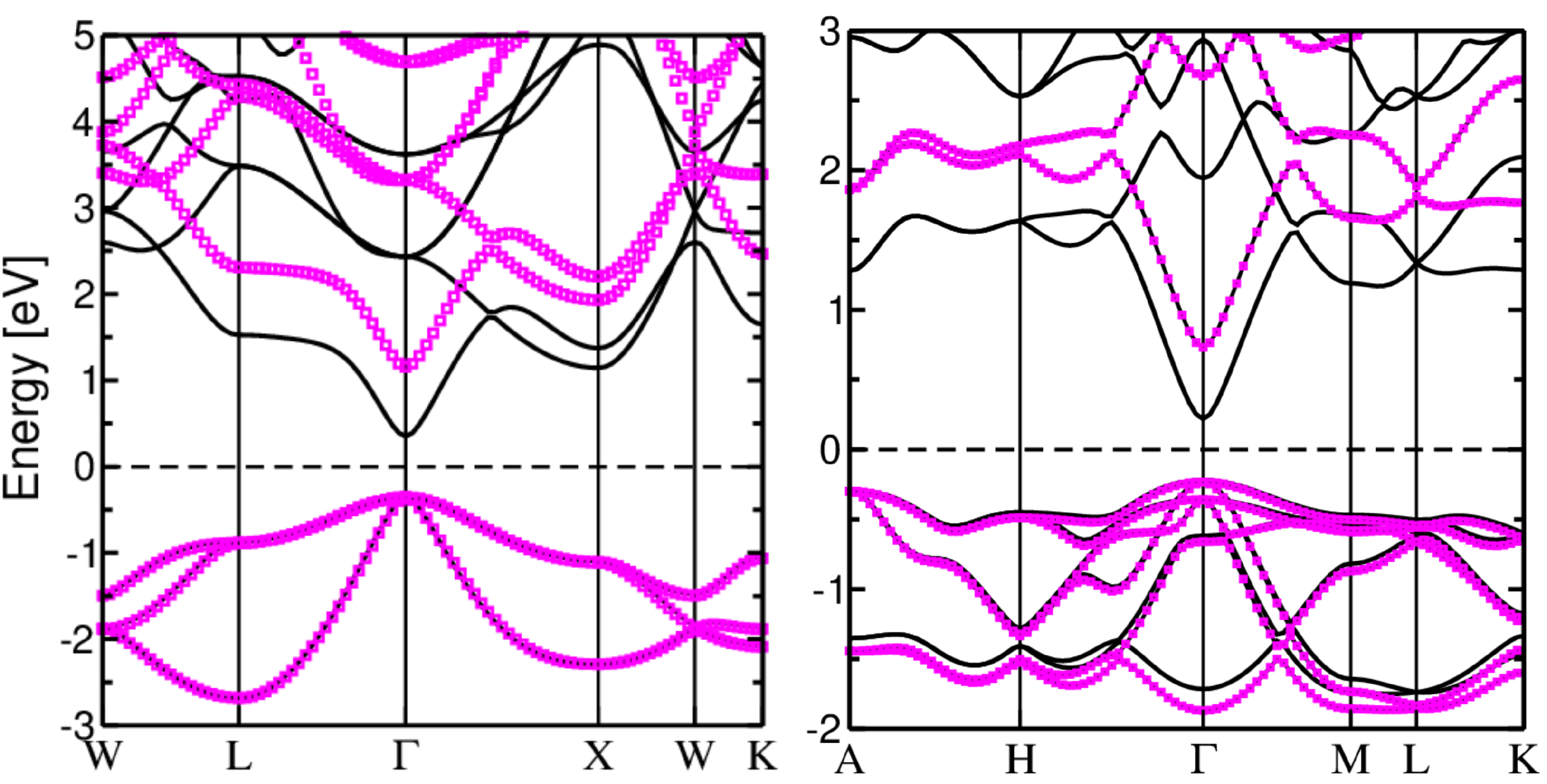}
	\caption{Band structure of Na$_2$KSb (left) and NaK$_2$Sb (right) computed from non-relativistic DFT (black lines) and $G_0W_0$ (magenta diamonds). The Fermi energy, marked by a dashed line, is set to zero in the midgap of the DFT result. DFT and $G_0W_0$ band structures are aligned at the VBM.}
	\label{fig:GWBS}
\end{figure}

In the next step of our investigation, we focus on the electronic structure of the materials.
Both Na$_2$KSb and NaK$_2$Sb, are characterized by a direct band gap at $\Gamma$ (see Fig.~\ref{fig:GWBS}).
As expected, the results provided by DFT with the semi-local PBE functional (Fig.~\ref{fig:GWBS}, black curves) severely underestimate the size of the gap compared to $G_0W_0$ (magenta diamonds).
The inclusion of the QP correction leads to a band-gap of 1.51~eV in Na$_2$KSb (Fig.~\ref{fig:GWBS}, left) and of 0.96~eV in NaK$_2$Sb (Fig.~\ref{fig:GWBS}, right).
Both values are more than twice as large as those yielded by DFT (0.70~eV in Na$_2$KSb and 0.44~eV in NaK$_2$Sb).
An early theoretical work based on the localized spherical wave method predicted for Na$_2$KSb a gap of the order of 1~eV~\cite{ette-degr00prb}.

By aligning the DFT and $G_0W_0$ band structures at the VBM, we are able to appreciate the overall effect of the QP correction in the valence and conduction regions (see Fig.~\ref{fig:GWBS}). 
In the case of Na$_2$KSb, the occupied states that are closest to the gap are identically reproduced by DFT and $G_0W_0$.
On the other hand, in the valence region of the band-structure of NaK$_2$Sb we notice a lowering by approximately 100 meV of the occupied  band around -1.5 eV, when the self-energy contribution is accounted for (see Fig.~\ref{fig:GWBS}, right).
This finding highlights the importance of performing a $G_0W_0$ calculation to solve the QP equation for all the electronic states of interest.
In the conduction region of both materials, the QP correction exerts an almost rigid shift to the bands. 
No significant variations affect the band curvatures and the effective masses.

\begin{figure*}
	\centering
	\includegraphics[width=0.85\textwidth]{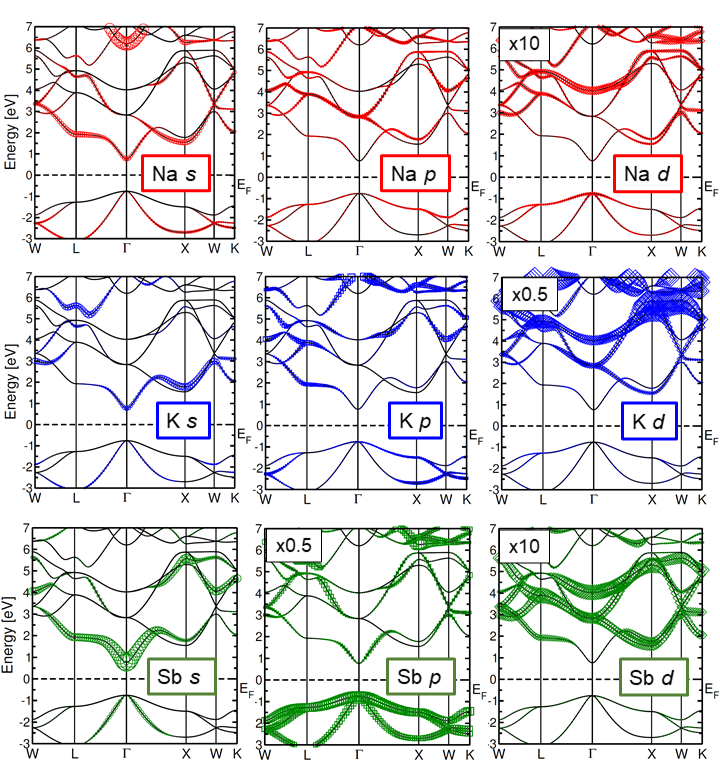}
	\caption{Quasi-particle band structure of cubic Na$_2$KSb with the orbital contributions of each species highlighted in color. The Fermi energy is set to zero in the mid-gap.}
	\label{fig:CubeBandStructure}
\end{figure*}

\begin{figure*}
	\centering
	\includegraphics[width=0.85\textwidth]{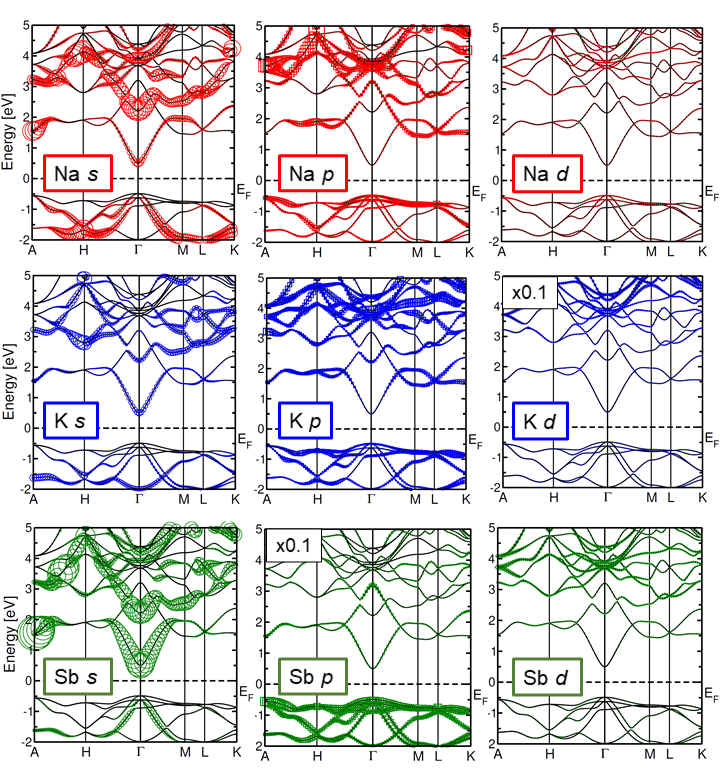}
	\caption{Quasi-particle band structure of hexagonal NaK$_2$Sb with the orbital contributions of each species highlighted in color. The Fermi energy is set to zero in the mid-gap.}
	\label{fig:HexBandStructure}
\end{figure*}

Further insight about the electronic properties can be gained by inspecting the orbital-projected character of the band structure (see Fig.~\ref{fig:CubeBandStructure} for cubic Na$_2$KSb and Fig.~\ref{fig:HexBandStructure} for hexagonal NaK$_2$Sb).
This analysis is relevant from many different perspectives: First, understanding the nature of the electronic states in the valence and conduction region provides the key ingredients to interpret and predict the optical absorption of the material.
Second, the non-stoichimetric compositions and/or the polycrystalline nature often exhibited by multi-alkali antimonide samples~\cite{schm+18prab} demand reliable theoretical references to identify specific structures or crystalline phases, for example, via x-ray spectroscopy~\cite{cocc+19sr}.
For this purpose, the knowledge of the atomic and orbital character of the bands is essential for the interpretations of the corresponding experiments~\cite{cocc-drax15prb,cocc+16prb,vorw+18jpcl}.
Finally, gaining information about the atomic and orbital character of the wave-functions is also essential in view of engineering multi-alkali antimonides compositions and alloys with enhanced performance in photocathode applications.

In both considered systems, the valence band is dominated by Sb $p$-states which are subject to a spin-orbit splitting of a few hundreds meV at the high-symmetry points in the Brillouin zone (see Fig.~\ref{fig:soc} and related discussion in the Appendix).
This leads to an overall reduction in the band-gap on the order of 0.2~eV.
These findings are in agreement with previous results obtained on the same materials~\cite{yala+18jmmm} as well as on related compounds~\cite{khan+20ijer,sass-cocc21}.
In the hexagonal phase, additional contributions from both Na and K $p$-orbitals, as far as the higher and flatter bands are concerned, and by $s$-states from Sb and Na in lower-energies and more dispersive bands.
The conduction band minimum (CBm), at the bottom of a parabolic band, has a marked $s$-like character in both materials, with dominant contributions from Sb and, to a minor extent, also to K and Na.
A careful inspection of the band structure of cubic Na$_2$KSb (Fig.~\ref{fig:CubeBandStructure}) reveals that the minimum at X, which belongs to a different band than the CBm, has a hybridized character involving mainly Na $s$-states as well as Sb and K $d$-states.
Higher-energy bands include also contributions from $p$-states of all the involved species, as well as K $d$-states (see Fig.~\ref{fig:CubeBandStructure}). 

The scenario is more faceted in the case of the hexagonal crystal NaK$_2$Sb.
In the corresponding conduction region, bands directly above the CBm bear mainly Sb and $s$-like contributions (see Fig.~\ref{fig:HexBandStructure}). 
Higher-energy bands in the cubic phase are dominated by K $d$-states. 
In hexagonal NaK$_2$Sb, on the other hand, hybridized $sp$-orbitals of the alkali atoms (Na and K) mainly contribute to the conduction bands above the CBm (see Fig.~\ref{fig:HexBandStructure}).
The differences in the unoccupied states of the two examined materials offers promising perspectives to use x-ray absorption spectroscopy to identify their fingerprints~\cite{cocc20pssrrl} in polycrystalline samples, where the two structures can coexist.

\subsection{Optical properties}
\label{subsec:opt}

\begin{figure}[h!]
	\centering
	\includegraphics[width=0.45\textwidth]{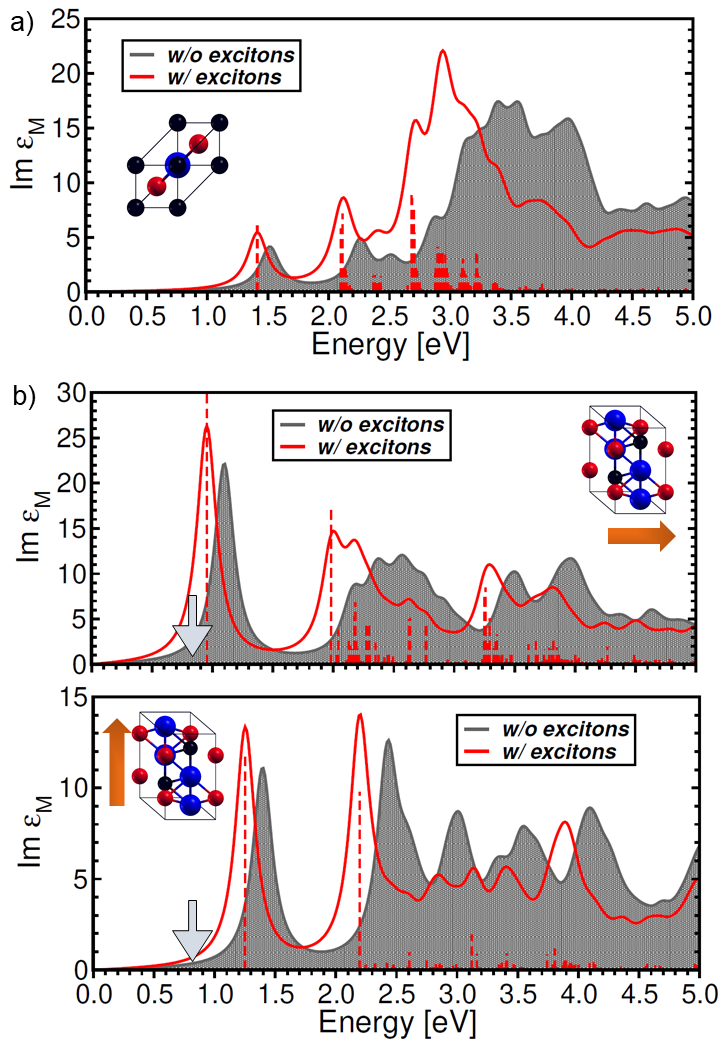}
	\caption{Imaginary part of the macroscopic dielectric functions of a) Na$_2$KSb and b) NaK$_2$Sb along the $a$ and $c$ axis of the crystal (panel . Red solid lines and grey shaded areas indicate results obtained with and without excitonic effects. Vertical dashed bars marking the position of the excitations, indicate to the discrete solutions of the BSE. The arrow in panel b) points to the lowest-energy dark excitation. A Lorentzian broadening of 100~meV is applied to all spectra.}
	\label{fig:spectra}
\end{figure}

The electronic structure discussed above is the cornerstone to interpret the imaginary parts of the dielectric functions calculated for the two materials (see Fig.~\ref{fig:spectra}).
This quantity provides information about the optical absorption of the systems.
In both Na$_2$KSb and in NaK$_2$Sb, the first bright excitation is found at the boundary between the visible and the infrared region. 
In the case of cubic Na$_2$KSb (Fig.~\ref{fig:spectra}a), where the dielectric function is fully defined by one diagonal component, the first maximum corresponding to the first excited state appears at 1.41~eV.
This peak is formed by one single excitation, which is triple-degenerate for symmetry reasons, and has a binding energy of 100~meV.
The latter quantity is estimated as the difference between the peak energy in the BSE spectrum and the QP gap, which also coincides with the energy of the first peak in the spectrum computed without excitons (independent QP approximation, IQPA, shown in  Fig.~\ref{fig:spectra}a).
In spite of its peaked shape, the first maximum in Fig.~\ref{fig:spectra}a) is not excitonic in nature, as it is present in the same form and with almost unaltered strength also in the IQPA spectrum (shaded grey area).
From the analysis of the BSE eigenvectors, computed from Eq.~\ref{bsehamiltonian}, we find that this excitation corresponds to the vertical transition at $\Gamma$ between the Sb $p$-like VBM and the Sb $s$-like CBm.
Above the first maximum, a second peak is found at 2~eV, followed by additional ones between 2.5 and 3.5~eV.
These higher-energy excitations experience a more considerable redistribution of the oscillator strength compared to the first one, which is ascribed to excitonic effects.
Our results are in excellent agreement with the available experimental reference~\cite{ebin-taka73prb}, where the maximum at 3~eV is particularly evident but also the absorption onset at about 1.5~eV and a shoulder close to 2~eV appears.

\begin{figure}[h!]
	\centering
	\includegraphics[width=0.48\textwidth]{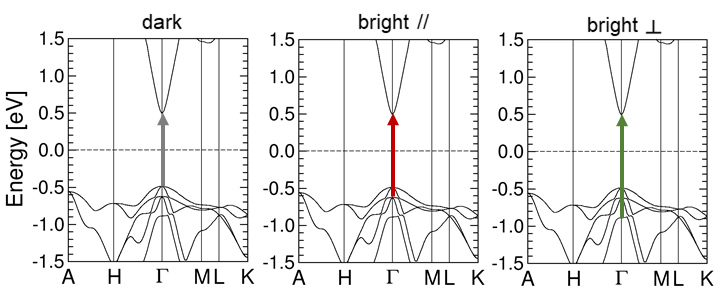}
	\caption{Schematic representation of the main single-particle contributions to the lowest-energy excitations in hexagonal NaK$_2$Sb, including the double-degenerate dark excitations at lowest energy (left) as well as the first bright peak in the in-plane (middle) and out-of-plane component (right) of the spectrum shown in Fig~\ref{fig:spectra}b).}
	\label{fig:exciton-hex}
\end{figure}

The dielectric function of hexagonal NaK$_2$Sb is given by two independent components, corresponding to the \textit{in-plane} and the \textit{out-of-plane} response of the material (see Fig.~\ref{fig:spectra}b and insets).
Both components are characterized by an intense peak at lowest-energy, namely at 0.95~eV and 1.25~eV in the in-plane and in the out-of-plane direction, respectively.
In both cases, an optically silent excitation is present at lowest energy (0.82~eV), stemming from the transition at $\Gamma$ between the first and second uppermost valence bands and the CBm (see Fig.~\ref{fig:exciton-hex}, left panel).
These valence states are those where the Na and K $p$-character is maximized (Fig.~\ref{fig:HexBandStructure}): the transition probability to the Sb $s$-like CBm is reasonably minimized.
Similar to the cubic phase and in spite of their sharp character, also in hexagonal NaK$_2$Sb, the first peaks are not excitonic, as a maximum of comparable intensity is found also in the corresponding IQPA spectra (see Fig.~\ref{fig:spectra}b). 
Due to the presence of a dark excitation at lowest energy, we adopt in this case an alternative and more appropriate definition of the exciton binding energy, given by the difference between the first maxima in the BSE and in the IQPA.
This definition is best applied to organic systems, where single-particle excitations are well distinguished from one another~\cite{cocc-drax15prb,cocc-drax17jpcm,beye+19cm}, as well as in composite systems, such as hybrid systems~\cite{fu+17pccp,vorw+18jpcl,turk+19ats} and heterostructures~\cite{aggo+17jpcl,lau+19prm,aggo+20prm}.
The binding energy associated to the lowest-energy peaks in the in-plane and out-of-plane optical components of NaK$_2$Sb is 150~meV.
Also the bright excitations at the two onsets stem from vertical transitions targeting the CBm at $\Gamma$.
The first bright excitation in the in-plane component of the spectrum, which is also double degenerate for symmetry reasons (one BSE solution is along the $a$-axis and the other one along $b$), comes from the third and fourth highest occupied bands (Fig.~\ref{fig:exciton-hex}, middle panel), which have a predominant Sb $p$-nature (see Fig.~\ref{fig:HexBandStructure}). 
Finally, first maximum in the out-of-plane component, is generated by a non-degenerate excitation from the parabolic valence band with Sb $sp$-character to the Sb $s$-like CBm (see Fig.~\ref{fig:exciton-hex}, right panel, and Fig.~\ref{fig:HexBandStructure}).

The higher-energy maxima in both components of the optical spectra of NaK$_2$Sb appear in the middle of the visible band, around 2~eV.
Interestingly, in the out-of-plane component, another sharp peak appears at 2.2~eV, again with a clear counterpart in the IQPA spectrum approximately 200~meV above it, which remarks its non-excitonic character.
Additional peaks are distinguishable at higher energies. 
Compared to the cubic material discussed above, in hexagonal NaK$_2$Sb, a redistribution of the oscillator strength to lower energies is still visible by comparing results of calculations with and without excitonic effects.
However, in this case, they mainly result in a rigid red-shift of the peaks.

\begin{figure*}
\begin{center}
\includegraphics[scale=1]{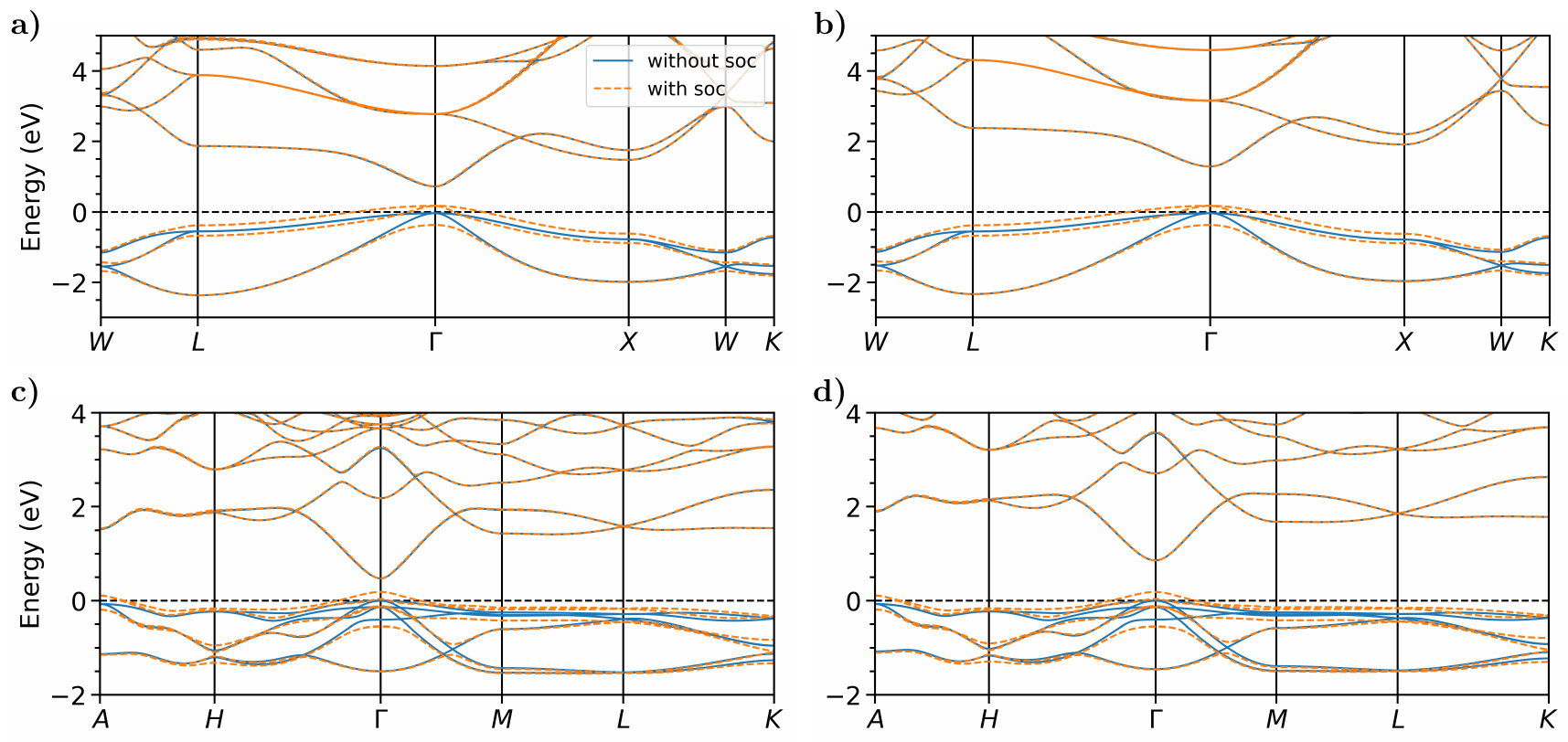}
\end{center}
\caption{Band structure of the gap region of Na$_2$KSb (top) and NaK$_2$Sb (bottom) calculated with and without spin-orbit coupling (SOC) on top of the PBE (a and c) and the SCAN functional (b and d). In all plots, the zero is set at the VBM calculated without SOC to enhance the visibility of the band splitting.}
\label{fig:soc}
\end{figure*}
\section{Summary and Conclusions}
\label{sec:conclu}
In summary, we have presented an \textit{ab initio} many-body study of the electronic and optical properties of Na$_2$KSb and NaK$_2$Sb, two members of the multi-alkali antimonide family of materials, which are regarded with interest for applications as vacuum electron sources.
Both materials exhibit a direct band gap at $\Gamma$, which is on the order of 1.5 eV in Na$_2$KSb and 1.0~eV in NaK$_2$Sb.
Spin-orbit coupling induces a reduction of the gap by approximately 200~meV in both systems.
The valence region is dominated by Sb $p$-states while the lowest conduction band has a pronounced parabolic dispersion due to the Sb $s$ orbital partially hybridized with the $s$-states of the alkali species too. 
Higher-energy unoccupied bands receive contributions from the $d$-electrons of all the atoms, with a predominance from potassium.
While these main characteristics are common to both crystals, fine details in the orbital hybridization in the conduction region, suggests the use of x-ray absorption spectroscopy to identify the two compounds in polycrystalline samples. 

The imaginary part of the dielectric functions of both Na$_2$KSb and NaK$_2$Sb are characterized by maxima at the boundary between the visible and the infrared region. 
The lowest-energy excitation in cubic Na$_2$KSb appears at 1.41~eV, approximately 0.5~eV higher in energy compared to the first bright peak in the dielectric function of NaK$_2$Sb.
In this material, a dark excitation is found at lowest energy (0.82~eV). 
In both materials, excitonic effects are marginal and mainly consist of a red-shift of the spectral weight towards lower energies.
Binding energies of the order of 100~meV are found for all the considered excitons.
Interestingly, the lowest-energy excitations are single-particle in nature, in spite of their relatively large oscillator strength and their sharply peaked shape. 

In conclusion, the electronic and optical properties computed for Na$_2$KSb and NaK$_2$Sb at the level of all-electron DFT and MBPT reveal desirable characteristics for the use of these materials as next-generation vacuum electron sources.
The absorption threshold of both materials around 1 -- 1.5~eV as well as the relatively weak binding strength of the lowest-energy excitations are particularly promising in view of photoemission in the infrared region.
The examined Na-based alkali antimonides exhibit similar electronic and optical properties as their Cs-based counterparts obtained within the same methodological framework~\cite{cocc+18jpcm,cocc+19sr}.
This finding hints that the replacement of the heavier Cs atom with the lighter Na does not affect the favorable photoemission characteristics of multi-alkali antimonides.
Even more importantly, this result suggests the potential for fine-tuning the electronic properties of these materials via tailored compositions and alloys. 

\section*{Data availability statement}
The data that support the findings of this study are available upon reasonable request from the authors.
\begin{acknowledgments}
The authors are grateful to Thorsten Kamps, Julius K{\"u}hn, Sonal Mistry, and Martin Schmei{\ss}er, for stimulating discussions. This work is partly funded by the German Federal Ministry of Education and Research (Professorinnenprogramm III) as well as from the Lower Saxony State (Professorinnen für Niedersachsen).
\end{acknowledgments}
\section*{Appendix: Effects of spin-orbit coupling on DFT}

In Fig.~\ref{fig:soc} we report the results of DFT calculations performed on Na$_\mathrm{2}$KSb and NaK$_\mathrm{2}$Sb including spin-orbit coupling. 
For this purpose, we used the all-electron, full potential code FHI-aims \cite{blum+09cpc} with the \textit{intermediate} default settings for the basis-set and the integration grids.
For Na$_\mathrm{2}$KSb and NaK$_\mathrm{2}$Sb, a 18$\times$18$\times$18 and a 12$\times$12$\times$6 \textbf{k}-meshes are used to sample the Brillouin zone.
The effect of spin-orbit coupling is accounted for based on a non-self-consistent post-self-consistent-field approximation \cite{huhn-blum17prm}.
For these calculations, we employ both PBE as well as the SCAN functional~\cite{sun+15prl}, implementing the meta-generalized-gradient-approximation, which was recently established to offer an optimal trade-off between accuracy and numerical costs for the evaluation of the band-gaps of Cs-based alkali-antimonides and tellurides~\cite{sass-cocc21}.

The impact of spin-orbit coupling is visible only in the valence region and in the vicinity of the high-symmetry points where bands are split (see Fig.~\ref{fig:soc}).
In Na$_\mathrm{2}$KSb, the magnitude of the band splitting ranges from 0.25~eV at the $W$-point up to about 0.55~eV at $\Gamma$.
No spitting occurs at the $K$-point.
In NaK$_\mathrm{2}$Sb, spin-orbit coupling splits the bands at all high-symmetry points, ranging from 0.15~eV at $H$ to 0.30~eV at $A$.
In both materials, the valence band maximum is shifted upwards at $\Gamma$, thereby reducing the band gap by about 0.2~eV.

Comparing now the effect of the xc functional, we notice that the most striking difference is the rigid shift of the conduction bands, which results in an increased band gap in both materials.
Using the SCAN functional we obtain a band gap of 1.32~eV and 0.86~eV for Na$_\mathrm{2}$KSb and NaK$_\mathrm{2}$Sb respectively.
These values are quite close to those calculated from $G_0W_0$ on top of PBE, in agreement with the results obtained for Cs$_\mathrm{3}$Sb at the same level of theory~\cite{sass-cocc21}.

%


\end{document}